*Review*

# Advanced Photocatalysts for CO₂ Conversion by Severe Plastic Deformation (SPD)

**Saeid Akrami [1,\*], Tatsumi Ishihara [2,3,4], Masayoshi Fuji [1,5] and Kaveh Edalati [2,3,\*]**

[1] Department of Life Science and Applied Chemistry, Nagoya Institute of Technology, Tajimi 507-0071, Japan; fuji@nitech.ac.jp

[2] WPI International Institute for Carbon-Neutral Energy Research (WPI-I2CNER), Kyushu University, Fukuoka 819-0395, Japan ishihara@cstf.kyushu-u.ac.jp

[3] Mitsui Chemicals, Inc.—Carbon Neutral Research Center (MCI-CNRC), Kyushu University, Fukuoka 819-0395, Japan

[4] Department of Applied Chemistry, Faculty of Engineering, Kyushu University, Fukuoka 819-0395, Japan

[5] Advanced Ceramics Research Center, Nagoya Institute of Technology, Tajimi 507-0071, Japan

[\*] Correspondence: saeidakrami91@gmail.com (S.A.); kaveh.edalati@kyudai.jp (K.E.); Tel.: +81-572-24-8110 (S.A.); +81-82-802-6744 (K.E.)

**Abstract:** Excessive CO₂ emission from fossil fuel usage has resulted in global warming and environmental crises. To solve this problem, the photocatalytic conversion of CO₂ to CO or useful components is a new strategy that has received significant attention. The main challenge in this regard is exploring photocatalysts with high efficiency for CO₂ photoreduction. Severe plastic deformation (SPD) through the high-pressure torsion (HPT) process has been effectively used in recent years to develop new active catalysts for CO₂ conversion. These active photocatalysts have been designed based on four main strategies: (i) oxygen vacancy and strain engineering, (ii) stabilization of high-pressure phases, (iii) synthesis of defective high-entropy oxides, and (iv) synthesis of low-bandgap high-entropy oxynitrides. These strategies can enhance the photocatalytic efficiency compared with conventional and benchmark photocatalysts by improving CO₂ adsorption, increasing light absorbance, aligning the band structure, narrowing the bandgap, accelerating the charge carrier migration, suppressing the recombination rate of electrons and holes, and providing active sites for photocatalytic reactions. This article reviews recent progress in the application of SPD to develop functional ceramics for photocatalytic CO₂ conversion.

## 1. Introduction

Nowadays, environmental crises, especially global warming caused by CO₂ emission from burning fossil fuels and humankind activities, are considered one of the most significant challenges in the world. Reduction of CO₂ to reactive CO gas or useful components and fuels, such as CH₄ and CH₃OH, using photocatalysts is one of the clean and new strategies, which is developing rapidly [1–3]. In photocatalytic CO₂ conversion, excited electrons transfer from the valence band to the conduction band of the photocatalysts by solar irradiation and contribute to the reduction of CO₂ to form desirable products, as shown in Figure 1a [3]. To perform these reduction reactions, some thermodynamic and kinetic conditions should be provided. From the viewpoint of thermodynamics, the standard potential of the reduction and oxidation reactions should be between the valence band and the conduction band of the photocatalyst [3,4]. On the other hand, from the kinetic viewpoint, the electrons should absorb the light, transfer to the conduction band, migrate to the surface of the photocatalyst, and take part in the reactions before combining with the holes [3,4]. To satisfy these kinetic and thermodynamic conditions, a photocatalyst should have some features, including high light absorbance, appropriate band structure, low recombination rate of electrons and holes, easy migration of charge carriers, and high surface affinity to adsorb $CO_2$ with abundant active sites [3,4]. A combination of these thermodynamic and kinetic factors determines the speed of the reactions and the type of final products in photocatalysis.

Semiconductors, such as $TiO_2$ [5–7], g-$C_3N_4$ [8,9], and $BiVO_4$ [10–12], are typical photocatalysts that have been engineered by various strategies to enhance the catalytic efficiency for $CO_2$ conversion. Doping with impurities, such as nitrogen, phosphorous, copper, and palladium [13–15]; defect engineering [16,17]; strain engineering [18,19]; mesoporous structure production [20]; and heterojunction introduction [21,22] are some of the most promising strategies that have been used so far to improve the optical properties and catalytic activity of various photocatalysts. Among these strategies, doping with impurities is the most investigated and feasible method, but impurities can increase the recombination rate of electrons and holes [13–15]. Therefore, finding new strategies to improve the photocatalytic activity and suppress the recombination rate of electrons and holes is a key issue. In this regard, severe plastic deformation (SPD) through the high-pressure torsion (HPT) method, which is mainly used for nanostructuring of metallic materials, has been used as a new tool to develop active photocatalysts for water splitting [23–30], dye degradation [31–34], and especially $CO_2$ conversion [35–38]. This method not only does not increase the recombination rate of electrons and holes but also effectively suppresses it and improves some other optical properties. The SPD method has also been used effectively to synthesize new families of catalysts, such as high-pressure photocatalysts and high-entropy photocatalysts [23,27].

This article reviews recent publications on the impact of SPD through the HPT method on photocatalytic activity for $CO_2$ conversion. The four main strategies used for this purpose are discussed in detail: (i) oxygen vacancy and strain engineering, (ii) stabilization of high-pressure phases, (iii) synthesis of defective high-entropy oxides, and (iv) synthesis of low-bandgap high-entropy oxynitrides.

## 2. Influence of HPT on Photocatalytic $CO_2$ Conversion

HPT as an SPD method has been used since 1935 until now for grain refinement and the production of nanostructured materials. In addition to grain refinement, the introduction of various defects, such as vacancies and dislocations, is another feature of HPT, which resulta in the improvement of the functionality of materials proceeded by this method [39,40]. In the HPT method, both large shear strain and high pressure (in the range of several gigapascals) are simultaneously utilized to process or synthesize various ranges of materials [39,40]. Strain and pressure are applied to the material (disc or ring shape) using two anvils that rotate with respect to each other, as shown in Figure 1b [41]. Due to the high processing pressure in HPT, it is applicable to hard and less ductile materials, such as high-melting temperature metals (hafnium [42], molybdenum [43], and tungsten [44]), amorphous glasses [45,46], silicon-based semiconductors [47,48], and even ultrahard diamond [49,50]. Another advantage of HPT is its capacity to induce ultra-SPD (i.e., shear strains over 1000 for mechanical alloying) [51]. The inducing ultra-SPD [51] together with fast dynamic diffusion [52,53] introduces the HPT method as a unique path to mechanically synthesize new materials even from immiscible systems [54,55]. Due to these unique features of HPT, the method was even used for the process and synthesis of hard and brittle ceramics, but the number of publications on ceramics is quite limited despite the high potential of these materials for various applications [23–38,56–81]. Published studies regarding ceramics processed or synthesized by HPT are presented in Table 1, although there are other classic publications on HPT processing of ceramics mainly by physicists and geologists [40].

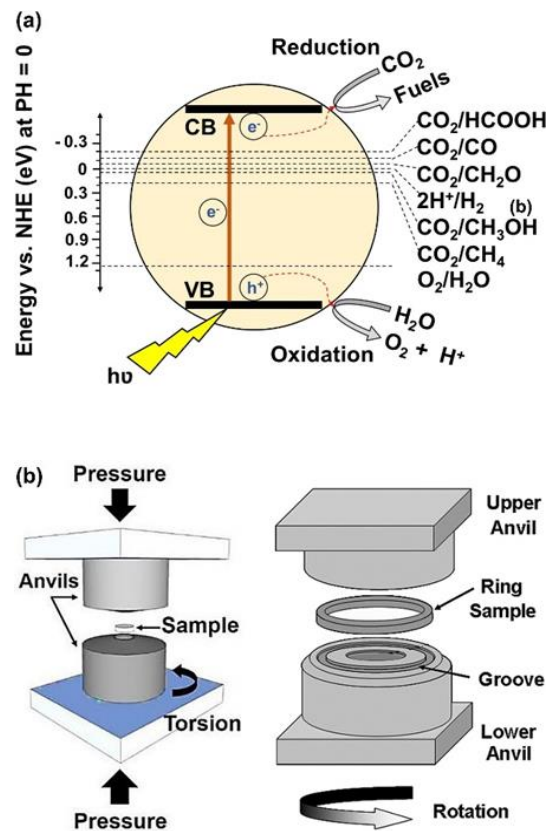

**Figure 1.** Schematic illustration of (**a**) photocatalytic $CO_2$ conversion and (**b**) high-pressure torsion [41].

**Table 1.** Summary of major publications about ceramics treated by high-pressure torsion and their major properties and applications.

| Materials | Investigated Properties and Applications | Reference |
|---|---|---|
| Various Materials | Impact of pressure and strain on allotropy | Bridgman (1935) [56] |
| $\alpha$-$Al_2O_3$ | Microstructure and mechanical properties | Edalati et al. (2010) [57] |
| $ZrO_2$ | Allotropic phase transformations | Edalati et al. (2011) [58] |
| CuO | Dielectric properties | Makhnev et al. (2011) [59] |
| CuO, $Y_3Fe_5O_{12}$, $FeBO_3$ | Optical properties and electronic structure | Gizhevskii et al. (2011) [60] |
| $ZrO_2$ | Phase transformation | Delogu et al. (2012) [61] |
| $Cu_2O$, CuO | Middle infrared absorption and X-ray absorption | Mostovshchikova et al. (2012) [62] |
| CuO, $Y_3Fe_5O_{12}$, $FeBO_3$ | Optical properties | Telegin et al. (2012) [63] |
| $BaTiO_3$ | Optical and dielectric properties | Edalati et al. (2015) [64] |
| $TiO_2$-II | Photocatalytic activity for hydrogen production | Razavi-Khosroshahi et al. (2016) [23] |
| Various Materials | Review on HPT | Edalati et al. (2016) [40] |
| $TiO_2$ | Plastic strain and phase transformation | Razavi-Khosroshahi et al. (2016) [65] |
| $Y_2O_3$ | Optical properties | Razavi-Khosroshahi et al. (2016) [66] |
| $YBa_2Cu_3O_y$ | Microstructural investigation | Kuznetsova et al. (2017) [67] |
| BN | Coupled elastoplasticity and plastic strain-induced phase transformation | Feng et al. (2017) [68] |

| ZnO | Photocatalytic activity for dye degradation | Razavi-Khosroshahi et al. (2017) [26] |
|---|---|---|
| $Fe_3O_4$ | Lithium-ion batteries | Qian et al. (2018) [69] |
| ZnO | Plastic flow and microstructural instabilities | Qi et al. (2018) [70] |
| $Fe_{71.2}Cr_{22.7}Mn_{1.3}N_{4.8}$ | Microstructural features | Shabashov et al. (2018) [71] |
| BN | Modeling of plastic flow and strain-induced phase transformation | Feng et al. (2019) [72] |
| $TiO_2$-II | Electrocatalysis for hydrogen generation | Edalati et al. (2019) [73] |
| $\gamma$-$Al_2O_3$ | Photocatalytic activity for dye degradation | Edalati et al. (2019) [27] |
| Various Oxides | Review on HPT of oxides | Edalati et al. (2019) [74] |
| MgO | Photocatalytic activity for dye degradation | Fujita et al. (2020) [28] |
| $ZrO_2$ | Photocatalytic activity for hydrogen production | Wang et al. (2020) [26] |
| $SiO_2$ | Photocatalytic activity for dye degradation | Wang et al. (2020) [34] |
| $CsTaO_3$, $LiTaO_3$ | Photocatalytic activity for hydrogen production | Edalati et al. (2020) [24] |
| GaN-ZnO | Photocatalytic activity for hydrogen production | Edalati et al. (2020) [25] |
| $Fe_{53.3}Ni_{26.5}B_{20.2}$, $Co_{28.2}Fe_{38.9}Cr_{15.4}Si_{0.3}B_{17.2}$ | Microstructure and mechanical properties | Permyakova et al. (2020) [75] |
| $TiHfZrNbTaO_{11}$ | Photocatalytic activity for hydrogen production | Edalati et al. (2020) [27] |
| $TiO_2$-ZnO | Photocatalytic activity for hydrogen production | Hidalgo-Jimeneza et al. (2020) [28] |
| $Bi_2O_3$ | Enhanced photocurrent generation | Fujita et al. (2020) [76] |
| $TiO_2$-II | Visible-light photocurrent generation | Wang et al. (2020) [77] |
| $TiO_2$-II | Photocatalytic activity for $CO_2$ conversion | Akrami et al. (2021) [30] |
| $TiZrHfNbTaO_6N_3$ | Photocatalytic activity for hydrogen production | Edalati et al. (2021) [29] |
| $SiO_2$, $VO_2$ | Phase transformation | Edalati et al. (2021) [78] |
| $TiO_2$ | Grain coarsening and phase transformation | Edalati et al. (2021) [79] |
| ZnO | Bandgap narrowing | Qi et al. (2021) [80] |
| $BiVO_4$ | Photocatalytic activity for $CO_2$ conversion | Akrami et al. (2022) [29] |
| $TiHfZrNbTaO_{11}$ | Photocatalytic activity for $CO_2$ conversion | Akrami et al. (2022) [31] |
| $TiZrNbTaWO_{12}$ | Photocatalytic activity for oxygen production | Edalati et al. (2022) [30] |
| $TiZrHfNbTaO_6N_3$ | Photocatalytic activity for $CO_2$ conversion | Akrami et al. (2022) [32] |

As given in Table 1, the recent usage of HPT to process and synthesize ceramics for photocatalysis, especially photocatalytic $CO_2$ conversion, has shown a high potential of this method for the enhancement of photocatalytic activity [35–38]. The HPT method effectively leads to increased efficiency by narrowing the bandgap, increasing the light absorbance, aligning the band structure, introducing the interphases and active sites for chemical adsorption and reaction, and accelerating the charge carrier migration [35–38]. While the HPT method can control all these features simultaneously by simple mechanical treatment, chemical methods are not usually able to improve all these features at the same time. The main drawbacks of the HPT method are the small quantity of the sample and the low specific surface area of the catalyst due to the high pressure and strain utilized. However, upscaling the HPT method and increasing the specific surface area by a post-HPT treatment are issues that can be addressed in the future. The improvement of features of photocatalysts by HPT has been achieved using four main strategies, including simultaneous strain and oxygen vacancy engineering, the introduction of high-pressure phases, the formation of defective high-entropy phases, and the production of low-bandgap high-entropy oxynitride phases. The responsibility of each mentioned strategy to improve the photocatalytic $CO_2$ conversion activity is discussed in detail as follows. It should be noted that all photocatalytic $CO_2$ conversion experiments on HPT-processed catalysts were performed in an aqueous liquid phase inside a quartz photoreactor with a continuous flow of CO into the liquid phase and $NaHCO_3$ as the sacrificial agent.

### 2.1. Simultaneous Strain and Oxygen Vacancy Engineering

Oxygen vacancy engineering is an effective method that has been used to improve photocatalytic $CO_2$ conversion. Oxygen vacancies increase the photocatalytic efficiency by increasing the light absorbance, accelerating the charge carrier separation, and enhancing the surface reactions [35,36]. Oxygen vacancies on the surface of the photocatalysts act as active sites to trap the electrons for various ranges of reduction reactions. It was also observed that oxygen vacancies have a significant role in adsorbing and activating the $CO_2$ molecules and increasing the local electronic density [35,36].

$BiVO_4$ is one of the common photocatalysts utilized for photocatalytic $CO_2$ conversion, but it suffers from a high recombination rate of electrons and holes and an inappropriate conduction band position [35]. Different strategies have been used to solve these problems, but in all of them, impurity atoms or a second phase are added to this material [35]. The HPT method was used to solve the problems of $BiVO_4$ for photocatalytic $CO_2$ conversion by simultaneous engineering of strain and oxygen vacancies without the addition of impurities. $BiVO_4$ was processed by HPT for $N$ = 0.25, 1, and 4 turns to investigate the impact of strain on photocatalytic properties and efficiency. Increasing the lattice strain and decreasing the crystallite size by increasing the HPT turns is shown in Figure 2a. The occurrence of lattice strain was also confirmed by Raman peak shift to lower wavenumbers, as shown in Figure 2b. It was also observed that the concentration of oxygen vacancies increases in $BiVO_4$ by increasing the applied shear strain. Figure 2c illustrates the oxygen vacancy concentration, calculated by X-ray photoelectron spectroscopy (XPS), against the number of HPT turns, confirming that the concentration of vacancies increases by increasing the applied shear strain. Furthermore, strain and vacancy engineering led to an increase in light absorbance (Figure 2d) and a slight narrowing of the bandgap from 2.4 eV for the initial powder to 2.1 eV for the sample proceeded by HPT for $N$ = 4 turns [35].

Simultaneous strain and oxygen vacancy engineering could significantly solve the problem of $BiVO_4$ in terms of the high recombination rate of electrons and holes, as shown in Figure 2e. This figure demonstrates that the HPT method decreases the photoluminescence intensity, which is a piece of evidence for the suppression of recombination. Finally, this strategy was successful in improving the photocatalytic activity of $BiVO_4$, as shown in Figure 2f. The CO production rate from $CO_2$ photoreduction was effectively increased by increasing the number of HPT turns. This study was the first successful work that used simultaneous strain and oxygen vacancy engineering to improve the photocatalytic activity of $BiVO_4$ without using impurities, suggesting SPD as a new path to improve the optical and electronic structure of photocatalysts for $CO_2$ conversion [35].

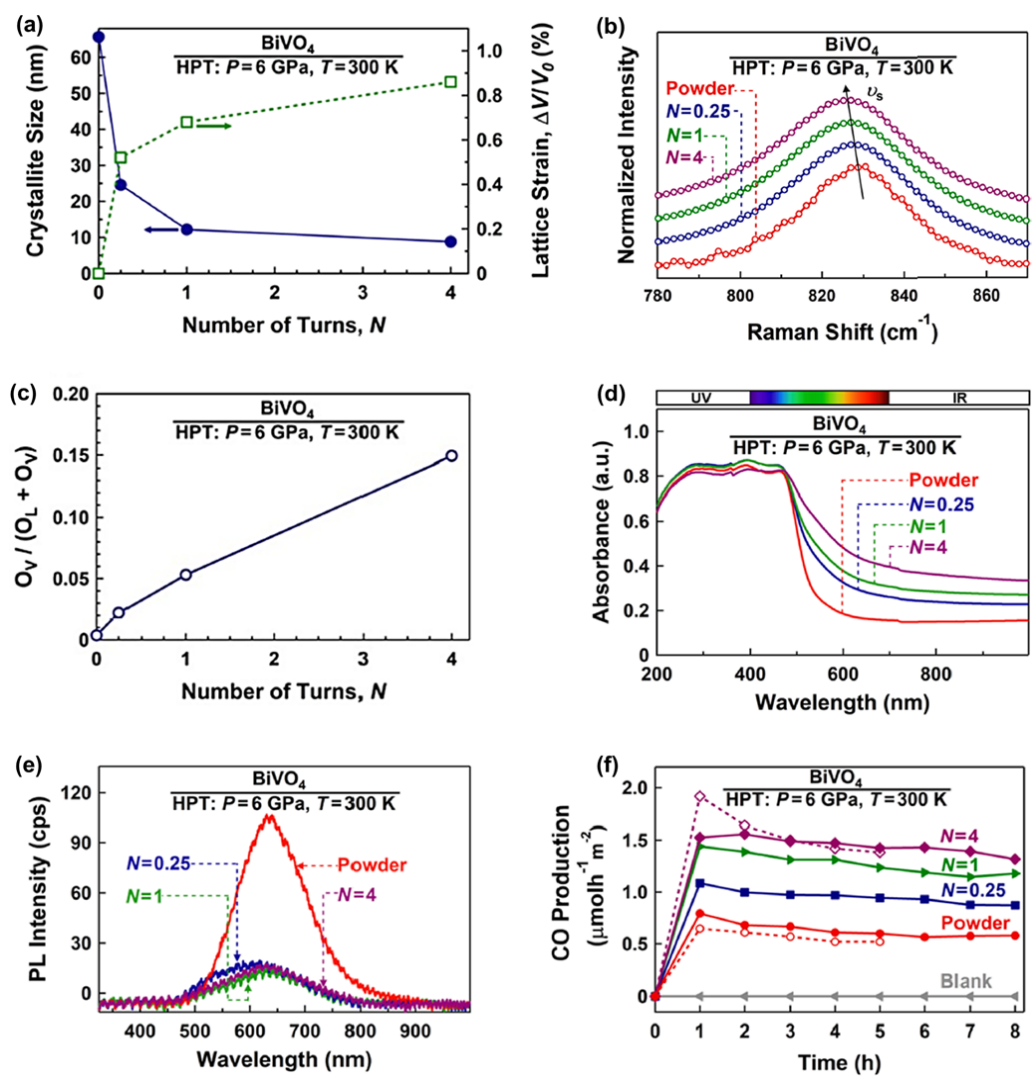

**Figure 2.** Improvement of light absorbance, suppression of recombination, and enhancement of photocatalytic $CO_2$ conversion for $BiVO_4$ by simultaneous strain and oxygen vacancy engineering using high-pressure torsion (HPT). (**a**) Crystallite size and volumetric strain versus the number of HPT turns ($N$), (**b**) Raman spectroscopy of initial and HPT-processed samples (inset: the appearance of samples), (**c**) oxygen vacancy concentration versus the number of HPT turns calculated by X-ray photoelectron spectroscopy, (**d**) UV–VIS spectroscopy, (**e**) photoluminescence spectra, and (**f**) photocatalytic CO production rate versus time for initial powder and sample proceeded by HPT for $N$ = 0.25, 1, and 4 turns [35].

## 2.2. Introducing High-Pressure Phases

The formation of high-pressure phases is one of the HPT effects that can occur for some ceramics, such as $TiO_2$ [65], $ZrO_2$ [58], ZnO [26], $SiO_2$ [34], $VO_2$ [78], $Y_2O_3$ [66], $BaTiO_3$ [64], $Al_2O_3$ [27], and BN [68]. It was observed that these high-pressure phases contain defects, such as oxygen vacancies and dislocations, and have nanosized grains, which makes them attractive for photocatalytic applications. $TiO_2$ with the anatase and rutile crystal structures is one of the most active photocatalysts for $CO_2$ conversion. As shown in Figure 3a, in addition to anatase and rutile, $TiO_2$ has a high-pressure $TiO_2$-II (columbite) phase with the orthorhombic structure. Despite many studies on $TiO_2$ photocatalysts, there was not any research work on photocatalytic $CO_2$ conversion on the $TiO_2$-II phase until 2021. Groups of current authors stabilized the $TiO_2$-II phase by the HPT method and investigated it for photocatalytic $CO_2$ conversion [36]. To decrease the fraction of oxygen vacancies in the bulk, which can act as recombination centers, an HPT-processed sample was further treated by annealing [36]. The formation of high-pressure $TiO_2$-II was proved by X-ray diffraction (XRD), Raman spectroscopy, and transmission electron microscopy (TEM). Raman spectra along with the appearance of samples are shown in Figure 3b. New Raman peaks at wavenumbers 171, 283, 316, 340, 357, 428, 533, and 572 cm⁻¹ correspond to the $TiO_2$-II phase. The changes in the color of the sample from white to dark green after HPT processing and from dark green to white after annealing indicate that large fractions of oxygen vacancies are formed after HPT processing, while some of them are annihilated after annealing, a fact that was also proved by various characterization techniques [36].

The light absorbance of the $TiO_2$-II phase produced by HPT processing was higher, and it had a narrower optical bandgap of 2.5 eV compared with anatase (3 eV), although the bandgap slightly increased to 2.7 eV after annealing [36]. Introducing the high-pressure $TiO_2$-II phase using HPT suppressed the recombination rate of electrons and holes. It also had a positive impact on photocurrent generation, as shown in Figure 3c so that the annealed sample had the highest current density, suggesting the improvement of charge carrier separations by introducing the high-pressure $TiO_2$-II phase. The potential of this new phase for $CO_2$ adsorption was measured by attenuated total reflectance Fourier transform infrared (ATR-FTIR) spectroscopy. It was observed that the annealed sample had the highest potential for $CO_2$ adsorption, which can help with photocatalytic $CO_2$ conversion. Finally, this new phase showed a higher potential for photocatalytic CO production compared with the anatase phase, as shown in Figure 3d. The introduction of the $TiO_2$-II phase with an optimized fraction of oxygen vacancies significantly improved the activity so that the annealed sample had the highest efficiency for $CO_2$-to-CO conversion. The formation of anatase–columbite interphases can also contribute to the high activity of the HPT-processed sample by increasing the electron–hole separation and migration. In conclusion, high-pressure phases show great potential to be used as photocatalysts, and SPD can be used to stabilize these high-pressure phases under ambient conditions [36].

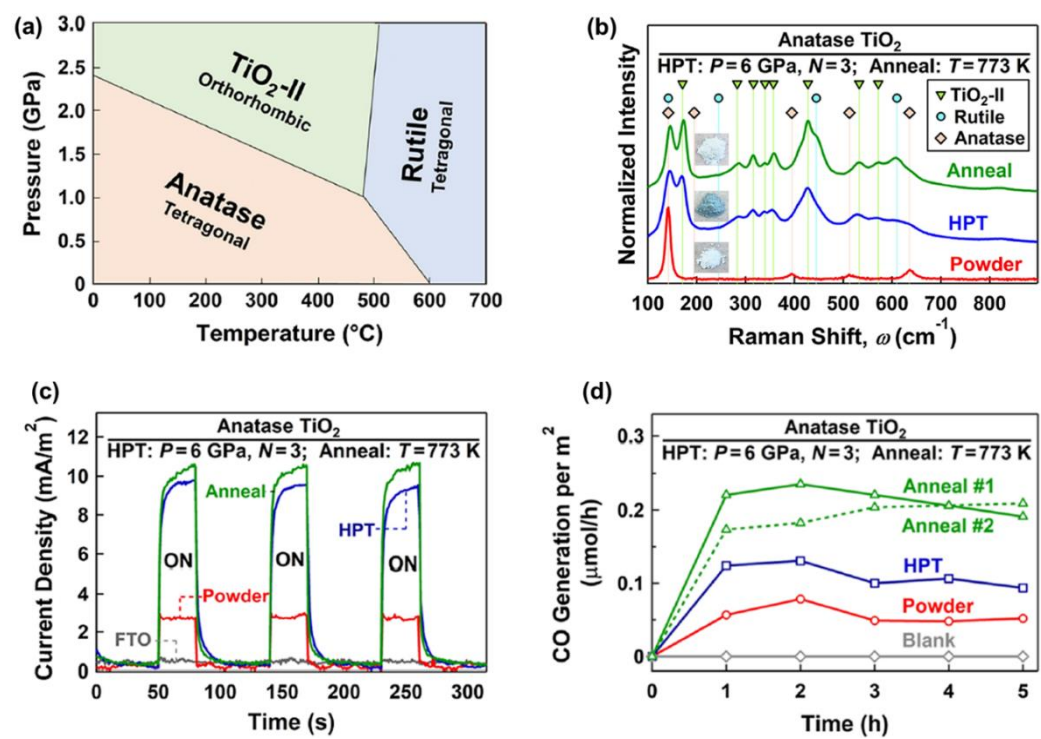

**Figure 3.** Improved charge carrier migration and photocatalytic $CO_2$ conversion by introducing the high-pressure $TiO_2$-II phase. (**a**) Pressure–temperature phase diagram of $TiO_2$. (**b**) Raman spectra, (**c**) photocurrent spectra, and (**d**) photocatalytic CO production rate versus time for $TiO_2$ before and after high-pressure torsion processing and after annealing [36].

## 2.3. Formation of Defective High-Entropy Phases

Introducing high-entropy ceramics as new materials with five or more principal elements opened a new path in the field of materials science to produce materials with high functionality for various applications [81,82]. High configurational entropy caused by a large number of elements in these materials leads to decreasing the Gibbs free energy and improving the phase stability. High-entropy ceramics have been utilized for various applications, and in many cases, they have shown better efficiencies than conventional materials [81,82]. Li-ion batteries [83], catalysts [84], dielectrics [85], magnetic components [86], thermal barrier coating [87], and so on are some of the applications of these materials. The high potential of high-entropy ceramics for various applications is attributed to their high stability, cocktail effect, lattice distortion, inherent defects, and valence electron distribution [81,82]. Despite the high functionality of these materials, their application for photocatalytic $CO_2$ conversion was not investigated until a study was conducted by the current authors in 2022 [37].

The HPT method, followed by oxidation, was used to fabricate a defective high-entropy oxide (HEO) with the composition of $TiZrNbHfTaO_{11}$ and dual crystal structure of monoclinic and orthorhombic [37]. The selection of elements for this high-entropy ceramic was conducted by considering the $d^0$ electronic structure of cations that have shown high potential for photocatalysis. The oxidation states of anionic and cationic elements and their uniform distribution were proved by XPS and energy-dispersive X-ray spectroscopy (EDS), respectively. The microstructure of the oxide is shown in Figure 4a using scanning electron microscopy (SEM) and in Figure 4b using high-resolution TEM. In addition to a nanocrystalline dual-phase structure, the material exhibited the presence of various defects, such as vacancies and dislocations, as shown in Figure 4b. The formation of oxygen vacancies in this material was examined by electron paramagnetic resonance (EPR) spectroscopy. These oxygen vacancies can act as shallow traps between the valence band and the conduction band for easier charge carrier separation, as shown in Figure 4c [37].

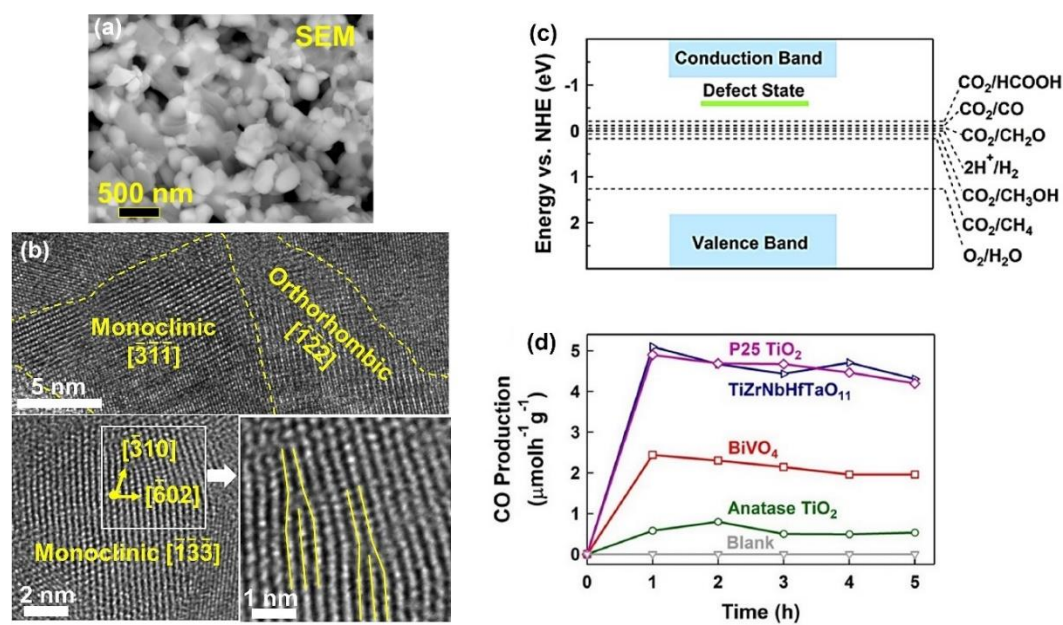

**Figure 4.** High photocatalytic $CO_2$ conversion on defective high-entropy oxide $TiZrNbHfTaO_{11}$ synthesized by high-pressure torsion. Microstructure of $TiZrNbHfTaO_{11}$ by (**a**) scanning electron microscopy and (**b**) high-resolution transmission electron microscopy. (**c**) Electronic band structure of $TiZrNbHfTaO_{11}$. (**d**) Photocatalytic CO production rate on $TiZrNbHfTaO_{11}$ versus time compared with P25 $TiO_2$, $BiVO_4$, and anatase $TiO_2$ [37].

$TiZrNbHfTaO_{11}$ had a higher light absorbance and lower bandgap compared with the binary oxides, including $TiO_2$, ZnO, $Nb_2O_5$, $HfO_3$, and $Ta_2O_5$ [37]. This HEO could successfully generate photocurrent, which indicates its potential for easy separation of electrons and holes to improve photocatalytic activity. $TiZrNbHfTaO_{11}$ showed higher activity for photocatalytic CO production compared with $BiVO_4$ and $TiO_2$ as two typical photocatalysts, as shown in Figure 4d. Additionally, it had the same photocatalytic activity as P25 $TiO_2$ as a benchmark photocatalyst. The high activity of this HEO for photocatalytic $CO_2$ conversion was attributed to the presence of defects, such as oxygen vacancies and dislocations; interphases; and high light absorbance. This study reported the first application of high-entropy ceramics for photocatalytic $CO_2$ conversion and introduced a new way to design and synthesize highly efficient high-entropy photocatalysts by SPD processing [37].

### *2.4. Synthesis of Low-Bandgap High-Entropy Oxynitrides*

Metal oxides are the most conventional photocatalysts for $CO_2$ conversion but suffer from a large bandgap. On the other hand, metal nitrides have a low bandgap but suffer from low stability compared with metal oxides. Metal oxynitrides are rather new materials that can solve the problems of metal oxides and nitrides in terms of large bandgap and low stability, respectively [88]. Although oxynitrides have been used for photocatalytic water splitting in many research works, their application for photocatalytic $CO_2$ conversion has been limitedly investigated mainly due to their limited chemical stability. The concept of high-entropy materials with high stability is one strategy used to produce high-entropy oxynitrides with low bandgap and high stability for $CO_2$ photoreduction [38].

A high-entropy oxynitride (HEON) with the composition of $TiZrNbHfTaO_6N_3$ was fabricated by the HPT method, followed by oxidation and nitriding, and its photocatalytic performance was compared with a corresponding HEO $TiZrNbHfTaO_6$ and P25 $TiO_2$ benchmark photocatalyst [38]. This HEON had dual phases with face-centered cubic (FCC) and monoclinic structures with uniform distribution of elements. This HEON material had much higher light absorbance compared with P25 $TiO_2$ and relevant HEO, as shown in Figure 5a. It showed a superior low bandgap of 1.6 eV as one of the lowest bandgaps reported in the literature for oxynitride photocatalysts. The improved electronic band structure of this HEON compared with P25 $TiO_2$ and HEO is shown in Figure 5b. The recombination rate of electrons and holes in HEON was low so that its photoluminescence intensity was negligible compared with P25 $TiO_2$ and HEO (Figure 5c). The shape of photocurrent spectra shown in Figure 5d also confirmed the low recombination rate of electrons and holes in this HEON compared with the HEO and P25 $TiO_2$ catalysts. The potential of this HEON for $CO_2$ adsorption was measured by diffuse reflectance infrared Fourier transform (DRIFT) spectrometry, which showed the higher physical adsorption and chemisorption (in the form of carbonate) of $CO_2$ on this HEON compared with P25 $TiO_2$ and HEO (Figure 5e).

This HEON successfully converted $CO_2$ to CO with extremely high efficiency even compared with the P25 $TiO_2$ benchmark photocatalyst, as shown in Figure 5f. Although HEON could adsorb the light in both visible and infrared regions of light, it could not convert $CO_2$ in these regions within the detection limits of the gas chromatograph. The stability of HEON was examined by conducting a long-term photocatalytic test for 20 h after storage of the sample in the air for 6 months. The photocatalytic activity of the material was not degraded, and X-ray diffraction analysis confirmed that the crystal structure of HEON did not change after 6-month storage and the long-time photocatalytic reaction. In conclusion, the low-bandgap HEON catalysts synthesized by SPD can be considered a new family of highly efficient photocatalysts for $CO_2$ conversion [38].

### 3. Discussion on Future Outlook

The application of SPD to synthesize new photocatalysts for $CO_2$ conversion introduced significant findings from the viewpoints of photocatalysis and SPD. The significance of these issues and their impact on the future outlook of this research field are discussed here.

For all these photocatalysts developed by HPT, CO was the only product that was detected using a flame ionization detector. The nonproduction of other products, such as $CH_4$, $CH_3OH$, HCOOH, or $CH_2=CH_2$, can be explained by considering the thermodynamic and kinetic parameters. For instance, $CH_4$ is a product that thermodynamically is more feasible to be produced than CO due to its lower standard potential. However, more electrons are required to produce this component compared with CO [89]. Therefore, from the viewpoint of kinetics, CO production is more feasible than $CH_4$ formation. Another point that should be considered is that CO has no tendency to be adsorbed to the active sites of the photocatalysts after production, which leads to propelling the reaction to CO production [89]. The production of CO as the only product can also be explained by the pathway of the reaction. In photocatalytic $CO_2$ conversion, the formation of a $CO_2^{\bullet-}$ intermediate product is the initial step. This intermediate product is formed by interchanging the electrons between $CO_2$ and the surface of the catalyst. Adsorption modes of $CO_2^{\bullet-}$ to the surface of the photocatalyst specify the reaction pathway. The $CO_2^{\bullet-}$ intermediate product can be adsorbed to the surface of the photocatalyst by three modes, which include (i) oxygen coordination, (ii) carbon coordination, and (iii) combination of oxygen and carbon coordination [90]. Oxygen coordination occurs when the photocatalyst is formed from tin, lead, mercury, indium, and cadmium metals. In this case, $^{\bullet}$OCHO and formic acid are produced as intermediate and final products, respectively. If the noble and transition metals are the elements forming the photocatalyst, carbon coordination occurs and $^{\bullet}$CO and CO are the intermediate and final products, respectively [90]. The presence of copper atoms in the structure of photocatalysts leads to the formation of a combination of oxygen and carbon coordination to produce both $^{\bullet}$OCHO and $^{\bullet}$CO as intermediates and CO, $CH_4$, and $C_2H_5OH$ as final products. Since all photocatalysts investigated by HPT include the transition metals, CO is the final product, and the reaction pathway can be considered as follows [90].

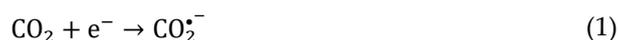

$$CO_2 + e^- \rightarrow CO_2^{\bullet-} \tag{1}$$

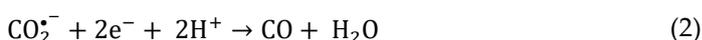

$$CO_2^{\bullet-} + 2e^- + 2H^+ \rightarrow CO + H_2O \tag{2}$$

Table 2 compares the photocatalytic activity of HEON synthesized by HPT with reported photocatalysts in the literature by normalizing the amount of CO production to catalyst mass and surface area [91–123]. Since the photocatalytic reaction occurs on the surface, comparing the results by normalizing them to the surface area is more reasonable. According to this table, the amount of CO production for HEON is $4.66 \pm 0.3$ μmolh$^{-1}$m$^{-1}$, which is higher than the best photocatalysts reported in the literature. This indicates that the contribution of SPD to introducing new families of photocatalysts will receive high appreciation in the future by considering the current demands in finding new strategies to deal with the $CO_2$ emissions; however, the synthesis method and compositions are expected to be modified by the experts in the field of photocatalysis. For example, one main disadvantage of SPD for the process and synthesis of catalysts is the low surface area of the synthesized material, while large specific surface areas are desirable in catalysis [74]. Moreover, theoretical studies are required to clarify the mechanisms underlying the high activity of photocatalysts developed by SPD so that new catalysts can be designed.

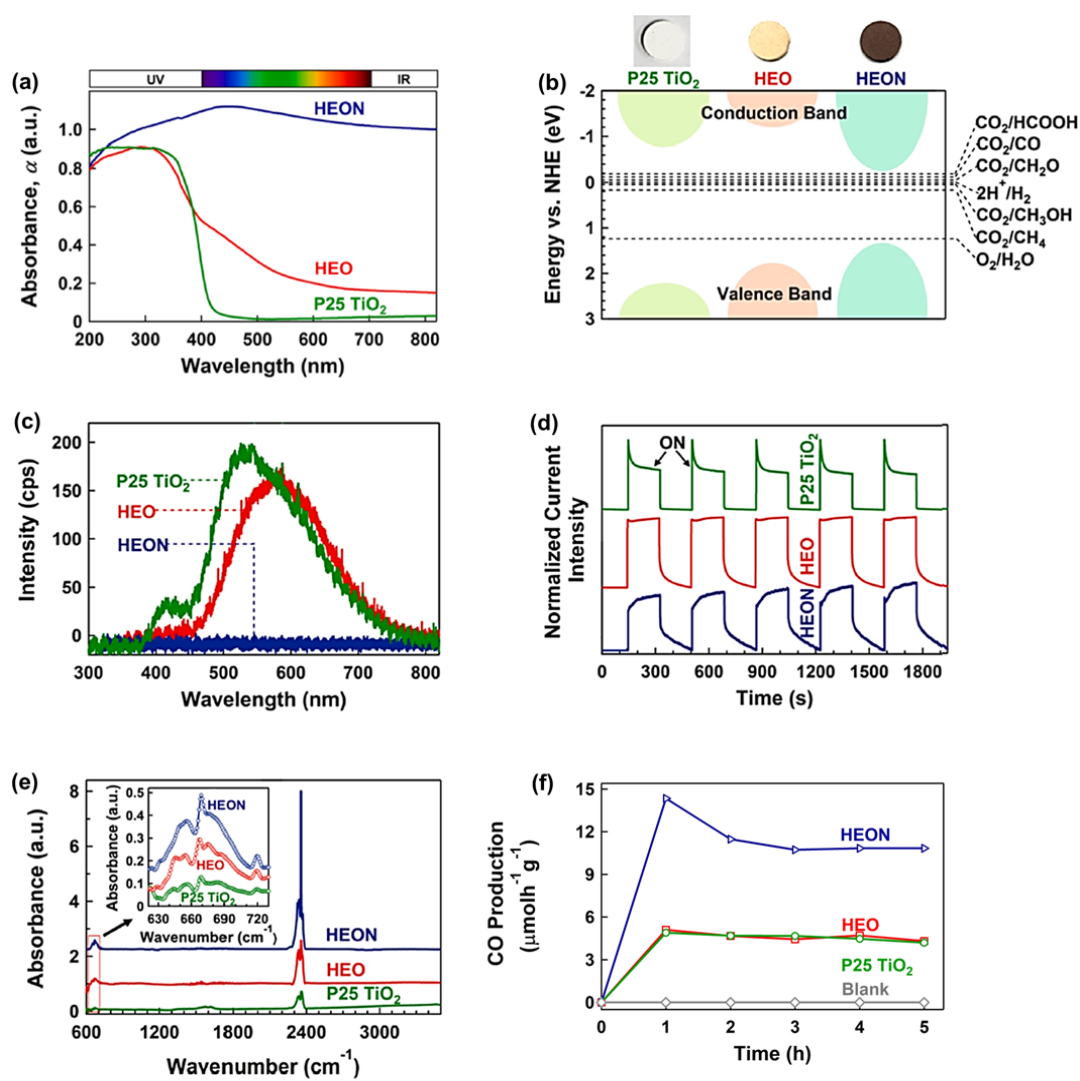

**Figure 5.** High light absorbance, appropriate band structure, suppressed recombination, significant $CO_2$ adsorption, and high photocatalytic $CO_2$ conversion for high-entropy oxynitride (HEON) $TiZrNbHfTaO_6N_3$. (**a**) UV–VIS light absorbance spectra, (**b**) electronic band structure together with the appearance of samples, (**c**) photoluminescence spectra, (**d**) photocurrent density versus time, (**e**) diffuse reflectance infrared Fourier transform spectra (peaks at 665 and 2350 $cm^{-1}$ represent chemisorption and physisorption of $CO_2$), and (**f**) photocatalytic CO production rate versus time for $TiZrNbHfTaO_6N_3$ compared with P25 $TiO_2$ and high-entropy oxide (HEO) $TiZrNbHfTaO_{11}$ [38].

The SPD field experienced significant progress in the past three decades, as discussed in several review papers [124–129], and more recently in a special issue in 2019 [130], which gathered overviews on both historical developments [131] and recent advancements [132]. A survey of these overviews indicates that despite significant progress on theoretical aspects [133,134], mechanisms [135,136], processing [137–144], microstructure [145–149], and mechanical properties [150–155] of metallic materials, there is a recent tendency to apply SPD to a wider range of materials (oxides [156], semiconductors [157], carbon polymorphs [158], glasses [159], and polymers [160]) to control phase transformations [161] and solid-state reaction [162–164] for achieving advanced functional properties [165–172]. $CO_2$ conversion is perhaps the newest application of SPD to functional materials, which expanded the synthesis capability of SPD from metallic materials to ceramics [37,38]. Moreover, this application has led to the introduction of new benchmark photocatalysts, which can open new pathways and research directions in corresponding fields. Although the application of SPD for $CO_2$ photoreduction is currently limited to the HPT method, which produces only small amounts of samples, the fundamentals developed by HPT should be used in the future to develop new methods with upscaled sample sizes and higher potential for industrial applications. This last issue is a general requirement of SPD for future commercialization in almost any application [173].

**Table 2.** Photocatalytic CO production rate on high-entropy oxynitride $TiZrNbHfTaO_6N_3$ synthesized by high-pressure torsion compared with photocatalysts reported in the literature.

| Photocatalyst | Catalyst Concentration | Light Source | CO Production Rate ($\mu molh^{-1}g^{-1}$) | CO Production Rate ($\mu molh^{-1}m^{-1}$) | Ref. |
|---|---|---|---|---|---|
| $TiO_2$/Graphitic Carbon | 100 mg (Gas System) | 300 W Xenon | 10.16 | 0.04 | Wang et al. (2013) [91] |
| Bicrystalline Anatase/Brookite $TiO_2$ Microspheres | 30 mg (Gas System) | 150 W Solar Simulator | 145 | 0.95 | Liu et al. (2013) [92] |
| Ag/TaON/RuBLRu′ | 2 gL⁻¹ (Liquid System) | 500 W High-Pressure Mercury | 0.056 | ---- | Sekizawa et al. (2013) [93] |
| 10 wt % Montmorillonite-Loaded $TiO_2$ | 50 mg (Gas System) | 500 W Mercury | 103 | 1.25 | Tahir et al. (2013) [94] |
| Anatase $TiO_2$ Nanofibers | 50 gL⁻¹ (Liquid System) | 500 W Mercury Flash | 40 | ----- | Zhang et al. (2013) [95] |
| $TiO_2$ Nanosheets Exposed {001} Facet | 1 gL⁻¹ (Liquid System) | Two 18 W Low-Pressure Mercury | 0.12 | 0.00095 | He et al. (2014) [96] |
| Anatase $TiO_2$ Hierarchical Microspheres | 200 mg (Gas System) | 40 W Mercury UV | 18.5 | 0.37 | Fang et al. (2014) [97] |
| $TiO_2$ and Zn(II) Porphyrin Mixed Phases | 60 mg (Gas System) | 300 W Xenon | 8 | 0.062 | Li et al. (2015) [98] |
| Anatase $TiO_2$ Hollow Sphere | 100 mg (Gas System) | 40 W Mercury UV | 14 | 0.16 | Fang et al. (2015) [99] |
| 10 wt % In-Doped Anatase $TiO_2$ | 250 mg (Gas System) | 500 W Mercury Flash | 81 | 1.33 | Tahir et al. (2015) [100] |
| $Pt^{2+}$–$Pt^0$/$TiO_2$ | 100 mg (Gas System) | 300 W Xenon | ~12.14 | 0.7 | Xiong et al. (2015) [101] |
| BiOI | 150 mg (Gas System) | 300 W High-Pressure Xenon | 4.1 | 0.03 | Ye et al. (2016) [102] |
| RuRu/Ag/TaON | 1 gL⁻¹ (Liquid System) | High-Pressure Mercury | 5 | ---- | Nakada et al. (2016) [103] |
| RuRu/TaON | 1 gL⁻¹ (Liquid System) | High-Pressure Mercury | 3.33 | ---- | Nakada et al. (2016) [103] |
| $CeO_{2-x}$ | 50 mg (Gas System) | 300 W Xenon | 1.65 | 0.08 | Ye et al. (2017) [104] |
| $Cu_2O$/$RuO_x$ | 500 mg (Gas System) | 150 W Xenon | 0.88 | --- | Pastor et al. (2017) [105] |
| $TiO_2$ 3D Ordered Microporous/Pd | 100 mg (Gas System) | 300 W Xenon | 3.9 | 0.066 | Jiao et al. (2017) [106] |
| $BiVO_4$/C/$Cu_2O$ | --- | 300 W Xenon | 3.01 | ---- | Kim et al. (2018) [107] |
| g-$C_3N_4$/$\alpha$-$Fe_2O_3$ | 200 mg (Gas System) | 300 W Xenon | 5.7 | ----- | Wang et al. (2018) [108] |
| x$Cu_2O$/$Zn_{2-2x}$Cr | 4 gL⁻¹ (Liquid System) | 200 W Mercury-Xenon | 2.5 | 0.018 | Jiang et al. (2018) [109] |

| TiO$_2$/Carbon Nitride Nanosheet | 25 mg (Gas System) | 150 W Xenon | 2.04 | ---- | Crake et al. (2019) [110] |
|---|---|---|---|---|---|
| TiO$_2$/CoO$_x$ Hydrogenated | 50 mg (Gas System) | 150 W UV | 1.24 | 0.0045 | Li et al. (2019) [111] |
| Bi$_4$O$_5$Br$_2$ | 20 mg (Gas System) | 300 W High-Pressure Xenon | 63.13 | 0.58 | Bai et al. (2019) [112] |
| ZnGaON | --- | 1600 W Xenon | 1.05 | --- | Maiti et al. (2019) [113] |
| C$_3$N$_4$ by Thermal Condensation | 100 mg (Gas System) | 350 W Mercury | 4.83 | ------ | Xia et al. (2019) [9] |
| Cd$_{1-x}$Zn$_x$S | 45 mg (Gas System) | UV-LED Irradiation | 2.9 | 0.015 | Kozlova et al. (2019) [114] |
| Bi$_{24}$O$_{31}$Cl$_{l10}$ | 50 mg (Gas System) | 300 W High-Pressure Xenon | 0.9 | --- | Jin et al. (2019) [115] |
| Bi$_2$Sn$_2$O$_7$ | 0.4 gL$^{-1}$ (Liquid System) | 300 W Xenon | 14.88 | 0.24 | Guo et al. (2020) [116] |
| Ag/Bi/BiVO$_4$ | 10 mg (Gas System) | 300 W Xenon Illuminator | 5.19 | 0.42 | Duan et al. (2020) [117] |
| g-C$_3$N$_4$/BiOCl | 20 mg (Gas System) | 300 W High-Pressure Xenon | 4.73 | --- | Chen et al. (2020) [118] |
| Fe/g-C$_3$N$_4$ | 1 gL$^{-1}$ (Liquid System | 300 W Xenon | ~22.5 | 0.06 | Dao et al. (2020) [119] |
| Bi$_2$MoO$_6$ | 0.7 gL$^{-1}$ (Liquid System) | 300 W Xenon | 41.5 | 1.26 | Zhang et al. (2020) [120] |
| g-C$_3$N$_4$/Zinc Carbodiimide/Zeolitic Imidazolate Framework | 100 mg (Gas System) | 300 W Xenon | ~0.45 | 0.014 | Xie et al. (2020) [121] |
| WO$_3$/LaTiO$_2$N | 10 mg (Gas System) | 300 W Xenon | 2.21 | 0.4 | Lin et al. (2021) [122] |
| α-Fe$_2$O$_3$/LaTiO$_2$N | 20 mg (Gas System) | 300 W Xenon | 9.7 | 0.65 | Song et al. (2021) [123] |
| TiZrHfNbTaO$_6$N$_3$ | 0.2 gL$^{-1}$ (Liquid System) | 400 W High-Pressure Mercury | 10.72 ± 1.77 | 4.66 ± 0.3 | Akrami et al. (2022) [32] |

## 4. Conclusions

Global warming has become a significant concern in recent years, which seriously threatens the life of creatures. Conversion of CO$_2$ molecules to other components, such as CO, is a way to stand this event. In this regard, photocatalytic CO$_2$ conversion, which uses solar irradiation as a clean energy source, has been introduced as a new and promising strategy in recent years. Despite the introduction of various materials, which are modified by various strategies, the efficiency of CO$_2$ photoreduction is still low compared with conventional methods for CO$_2$ conversion. High-pressure torsion (HPT) as a severe plastic deformation (SPD) method has been used recently to produce some of the new and most active photocatalysts for CO$_2$ conversion. The HPT method can increase the CO$_2$ photoreduction efficiency by (i) oxygen vacancy and strain engineering, (ii) the stabilization of high-pressure phases, (iii) the formation of defective high-entropy oxides, and (iv) the synthesis of low-bandgap oxynitrides.

**Author Contributions:** Conceptualization, S.A., T.I., M.F. and K.E.; writing—review and editing, S.A., T.I., M.F. and K.E. All authors have read and agreed to the published version of the manuscript.

**Funding:** The author S.A. thanks Hosokawa Powder Technology Foundation, Japan, for a grant. The author K.E. was supported in part by the MEXT, Japan, through Grants-in-Aid for Scientific Research (JP19H05176, JP21H00150, and JP22K18737), and in part by Mitsui Chemicals, Inc.

**Institutional Review Board Statement:** Not applicable.

**Informed Consent Statement:** Not applicable.

**Data Availability Statement:** No new data were created or analyzed in this study. Data sharing is not applicable to this article.

**Conflicts of Interest:** The authors declare no conflicts of interest.

## References

1. Forkel, M.; Carvalhais, N.; Rodenbeck, C.; Keeling, R.; Heimann, M.; Thonicke, K.; Zaehle, S.; Reichstein, M. Enhanced seasonal $CO_2$ exchange caused by amplified plant productivity in northern ecosystems. *Science* **2016**, *351*, 696–699.
2. Morris, A.J.; Meyer, G.J.; Fujita, E. Molecular approaches to the photocatalytic reduction of carbon dioxide for solar fuels. *Acc. Chem. Res.* **2009**, *42*, 1983–1994.
3. Tong, H.; Ouyang, S.; Bi, Y.; Umezawa, N.; Oshikiri, M.; Ye, J. Nano-photocatalytic materials: Possibilities and challenges. *Adv. Mater.* **2012**, *24*, 229–251.
4. Li, X.; Yu, J.; Jiang, C. Chapter 1—Principle and surface science of photocatalysis. *Interface Sci. Technol.* **2020**, *31*, 1–38.
5. Ola, O.; Maroto-Valer, M.M. Synthesis, characterization and visible light photocatalytic activity of metal based $TiO_2$ monoliths for $CO_2$ reduction. *Chem. Eng. J.* **2016**, *283*, 1244–1253.
6. Neatu, S.; Macia-Agullo, J.A.; Concepcion, P.; Garcia, H. Gold-copper nanoalloys supported on $TiO_2$ as photocatalysts for $CO_2$ reduction by water. *J. Am. Chem. Soc.* **2014**, *136*, 15969–15976.
7. Kocí, K.; Mateju, K.; Obalova, L.; Krejcikova, S.; Lacny, Z.; Placha, D.; Capek, L.; Hospodkova, A.; Solcova, O. Effect of silver doping on the $TiO_2$ for photocatalytic reduction of $CO_2$. *Appl. Catal. B* **2010**, *96*, 239–244.
8. Akhundi, A.; Habibi-Yangjeh, A.; Abitorabi, M.; Pouran, S.R. Review on photocatalytic conversion of carbon dioxide to value-added compounds and renewable fuels by graphitic carbon nitride-based photocatalysts. *Catal. Rev. Sci. Eng.* **2019**, *61*, 595–628.
9. Xia, P.; Antonietti, M.; Zhu, B.; Heil, T.; Yu, J.; Cao, S. Designing defective crystalline carbon nitride to enable selective $CO_2$ photoreduction in the gas phase. *Adv. Funct. Mater.* **2019**, *29*, 1900093.
10. Wang, K.; Zhang, L.; Su, Y.; Sun, S.; Wang, Q.; Wang, H.; Wang, W.:Boosted $CO_2$ photoreduction to methane via Co doping in bismuth vanadate atomic layers. *Catal. Sci. Technol.* **2018**, *8*, 3115–3122.
11. Huang, L.; Duan, Z.; Song, Y.; Li, Q.; Chen, L. $BiVO_4$ microplates with oxygen vacancies decorated with metallic Cu and Bi nanoparticles for $CO_2$ photoreduction. *ACS Appl. Nano Mater.* **2021**, *4*, 3576–3585.
12. Zhu, Z.; Yang, C.X.; Hwang, Y.T.; Lin, Y.C.; Wu, R.J. Fuel generation through photoreduction of $CO_2$ on novel $Cu/BiVO_4$. *Mater. Res. Bull.* **2020**, *130*, 110955.
13. Lee, G.J.; Anandan, S.; Masten, S.J.; Wu, J.J. Photocatalytic hydrogen evolution from water splitting using Cu doped ZnS microspheres under visible light irradiation. *Renew. Energy* **2016**, *89*, 18–26.
14. Liu, B.; Ye, L.; Wang, R.; Yang, J.; Zhang, Y.; Guan, R.; Tian, L.; Chen, X. Phosphorus-doped graphitic carbon nitride nanotubes with amino-rich surface for efficient $CO_2$ capture, enhanced photocatalytic activity, and product selectivity. *ACS Appl. Mater. Interfaces* **2018**, *10*, 4001–4009.
15. Kuvarega, A.T.; Krause, R.W.M.; Mamba, B.B. Nitrogen/palladium-codoped $TiO_2$ for efficient visible light photocatalytic dye degradation. *J. Phys. Chem. C* **2011**, *115*, 22110–22120.
16. Bai, S.; Zhang, N.; Gao, C.; Xiong, Y. Defect engineering in photocatalytic materials. *Nano Energy* **2018**, *53*, 293–336.
17. Di, J.; Zhu, C.; Ji, M.; Duan, M.; Long, R.; Yan, C.; Gu, K.; Xiong, J.; She, Y.; Xia, J.; et al. Defect-rich $Bi12O17Cl2$ nanotubes self-accelerating charge separation for boosting photocatalytic $CO_2$ reduction. *Angew. Chem.* **2018**, *57*, 14847–14851.
18. Cai, X.; Wang, F.; Wang, R.; Xi, Wang, Y.A.; Wang, J.; Teng, B.; Baiy, S. Synergism of surface strain and interfacial polarization on Pd@Au core-shell cocatalysts for highly efficient photocatalytic $CO_2$ reduction over $TiO_2$. *J. Mater. Chem. A* **2020**, *8*, 7350–7359.
19. Liu, Z.; Menendez, C.; Shenoy, J.; Hart, J.N.; Sorrell, C.C.; Cazorl, C. Strain engineering of oxide thin films for photocatalytic applications. *Nano Energy* **2020**, *72*, 104732.
20. Li, Y.; Wang, W.N.; Zhan, Z.; Woo, M.H.; Wu, C.Y.; Biswas, P. Photocatalytic reduction of $CO_2$ with $H_2O$ on mesoporous silica supported $Cu/TiO_2$ catalysts. *Appl. Catal. B* **2010**, *100*, 386–392.
21. Cao, S.; Shen, B.; Tong, T.; Fu, J.; Yu, J. 2D/2D heterojunction of ultrathin MXene/ $Bi_2WO_6$ nanosheets for improved photocatalytic $CO_2$ reduction. *Adv. Funct. Mater.* **2018**, *28*, 1800136.
22. Li, J.; Shao, W.; Geng, M.; Wan, S.; Ou, M.; Chen, Y. Combined Schottky junction and doping effect in $Cd_xZn_{1-x}S@Au/BiVO_4$ Z-Scheme photocatalyst with boosted carriers charge separation for $CO_2$ reduction by $H_2O$. *J. Colloid Interface Sci.* **2022**, *606*, 1469–1476.

23. Razavi-Khoroshahi, H.; Edalati, K.; Hirayama, M.; Emami, H.; Arita, M.; Yamauchi, M.; Hagiwara, H.; Ida, S.; Ishihara, T.; Akiba, E.; et al. Visible- light-driven photocatalytic hydrogen generation on nanosized TiO₂-II stabilized by high-pressure torsion. *ACS Catal.* **2016**, *6*, 5103–5107.

24. Edalati, K.; Fujiwara, K.; Takechi, S.; Wang, Q.; Arita, M.; Watanabe, M.; Sauvage, X.; Ishihara, T.; Horita, Z. Improved photocatalytic hydrogen evolution on tantalate perovskites CsTaO₃ and LiTaO₃ by strain-induced vacancies. *ACS Appl. Energy Mater.* **2020**, *3*, 1710–1718.

25. Edalati, K.; Uehiro, R.; Takechi, S.; Wang, Q.; Arita, M.; Watanabe, M.; Ishihara, T.; Horita, Z. Enhanced photocatalytic hydrogen production on GaN-ZnO oxynitride by introduction of strain-induced nitrogen vacancy complexes. *Acta Mater.* **2020**, *185*, 149–156.

26. Wang, Q.; Edalati, K.; Koganemaru, Y.; Nakamura, S.; Watanabe, M.; Ishihara, T.; Horita, Z. Photocatalytic hydrogen generation on low-bandgap black zirconia (ZrO₂) produced by high-pressure torsion. *J. Mater. Chem. A* **2020**, *8*, 3643–3650.

27. Edalati, P.; Wang, Q.; Razavi-Khoroshahi, H.; Fuji, M.; Ishihara, T.; Edalati, K.; Photocatalytic hydrogen evolution on a high entropy oxide. *J. Mater. Chem. A* **2020**, *8*, 3814–3821.

28. Hidalgo-Jimeneza, J.; Wang, Q.; Edalatib, K.; Cubero-Sesína, J.M.; Razavi-Khoroshahid, H.; Ikomac, Y.; Gutiérrez-Fallase, D.; Dittel-Mezaa, F.A.; Rodríguez-Rufinoa, J.C.; Fujid, M.; et al. Phase transformations, vacancy formation and variations of optical and photocatalytic properties in TiO₂-ZnO composites by high pressure torsion. *Int. J. Plast.* **2020**, *124*, 170–185.

29. Edalati, P.; Shen, X.F.; Watanabe, M.; Ishihara, T.; Arita, M.; Fuji, M.; Edalati, K. High-entropy oxynitride as a low-bandgap and stable photocatalyst for hydrogen production. *J. Mater. Chem. A* **2021**, *9*, 15076–15086.

30. Edalati, P.; Itagoe, Y.; Ishihara, H.; Ishihara, T.; Emami, H.; Arita, M.; Fuji, M.; Edalati, K. Visible-light photocatalytic oxygen production on a high-entropy oxide by multiple-heterojunction introduction. *J. Photochem. Photobiol. A* **2022**, *433*, 114167.

31. Razavi-Khoroshahi, H.; Edalati, K.; Wu, J.; Nakashima, Y.; Arita, M.; Ikoma, Y.; Sadakiyo, M.; Inagaki, Y.; Staykov, A.; Yamauchi, M.; et al. High-pressure zinc oxide phase as visible-light-active photocatalyst with narrow band gap. *J. Mater. Chem. A* **2017**, *5*, 20298–20303.

32. Edalati, K.; Fujita, I.; Takechi, S.; Nakashima, Y.; Kumano, K.; Razavi-Khoroshahi, H.; Arita, M.; Watanabe, M.; Sauvage, X.; Akbay, T.; et al. Photocatalytic activity of aluminum oxide by oxygen vacancy generation using high-pressure torsion straining. *Scr. Mater.* **2019**, *173*, 120–124.

33. Fujita, I.; Edalati, K.; Wang, Q.; Arita, M.; Watanabe, M.; Munetoh, S.; Ishihara, T.; Horita, Z. High-pressure torsion to induce oxygen vacancies in nanocrystals of magnesium oxide: Enhanced light absorbance, photocatalysis and significance in geology. *Materialia* **2020**, *11*, 100670.

34. Wang, Q.; Edalati, K.; Fujita, I.; Watanabe, M.; Ishihara, T.; Horita, Z. High-pressure torsion of SiO₂ quartz sand: Phase transformation, optical properties, and significance in geology. *J. Am. Ceram. Soc.* **2020**, *103*, 6594–6602.

35. Akrami, S.; Murakami, Y.; Watanabe, M.; Ishihara, T.; Arita, M.; Guo, Q.; Fuji, M.; Edalati, K. Enhanced CO₂ conversion on highly-strained and oxygen-deficient BiVO₄ photocatalyst. *Chem. Eng. J.* **2022**, *442*, 136209.

36. Akrami, S.; Watanabe, M.; Ling, T.H.; Ishihara, T.; Arita, M.; Fuji, M.; Edalati, K. High pressure TiO₂-II polymorph as an active photocatalyst for CO₂ to CO conversion. *Appl. Catal. B* **2021**, *298*, 120566.

37. Akrami, S.; Murakami, Y.; Watanabe, M.; Ishihara, T.; Arita, M.; Fuji, M.; Edalati, K. Defective high-entropy oxide photocatalyst with high activity for CO₂ conversion, *Appl. Catal. B* **2022**, *303*, 120896.

38. Akrami, S.; Edalati, P.; Shundo, Y.; Watanabe, M.; Ishihara, T.; Fuji, M.; Edalati, K. Significant CO₂ photoreduction on a high-entropy oxynitride, *Chem. Eng. J.* **2022**, *449*, 137800.

39. Edalati, K.; Bachmaier, A.; Beloshenko, V.A.; Beygelzimer, Y.; Blank, V.D.; Botta, W.J.; Bryła, K.; Čížek, J.; Divinski, S.; Enikeev, N.A.; et al. Nanomaterials by severe plastic deformation: Review of historical developments and recent advances. *Mater. Res. Lett.* **2022**, *10*, 163–256.

40. Edalati, K.; Horita, Z. A review on high-pressure torsion (HPT) from 1935 to 1988. *Mater. Sci. Eng. A* **2016**, *652*, 325–352.

41. Edalati, K.; Horita, Z. Scaling-up of high pressure torsion using ring shape. *Mater. Trans.* **2009**, *50*, 92–95.

42. Edalati, K.; Horita, Z.; Mine, Y. High-pressure torsion of hafnium. *Mater. Sci. Eng. A* **2010**, *527*, 2136–2141.

43. Lee, S.; Edalati, K.; Horita, Z. Microstructures and mechanical properties of pure V and Mo processed by high-pressure torsion. *Mater. Trans.* **2010**, *51*, 1072–1079.

44. Chen, L.; Ping, L.; Ye, T.; Lingfeng, L.; Kemin, X.; Meng, Z. Observations on the ductility and thermostability of tungsten processed from micropowder by improved high-pressure torsion. *Rare Met. Mater. Eng.* **2016**, *45*, 3089–3094.

45. Edalati, K.; Yokoyama, Y.; Horita, Z. High-pressure torsion of machining chips and bulk discs of amorphous Zr₅₀Cu₃₀Al₁₀Ni₁₀. *Mater. Trans.* **2010**, *51*, 23–26.

46. Wang, Y.B.; Qu, D.D.; Wang, X.H.; Cao, Y.; Liao, X.Z.; Kawasaki, M.; Ringer, S.P.; Shan, Z.W.; Langdon, T.G.; Shen, J. Introducing a strain-hardening capability to improve the ductility of bulk metallic glasses via severe plastic deformation. *Acta Mater.* **2012**, *60*, 253–260.

47. Edalati, K.; Horita, Z. Correlations between hardness and atomic bond parameters of pure metals and semi-metals after processing by high-pressure torsion. *Scr. Mater.* **2011**, *64*, 161–164.

48. Ikoma, Y.; Hayano, K.; Edalati, K.; Saito, K.; Guo, Q.; Horita, Z. Phase transformation and nanograin refinement of silicon by processing through high-pressure torsion. *Appl. Phys. Lett.* **2012**, *101*, 121908.

49. Blank, V.D.; Churkin, V.D.; Kulnitskiy, B.A.; Perezhogin, I.A.; Kirichenko, A.N.; Erohin, S.V.; Sorokin, P.B.; Popov, M.Y. Pressure-induced transformation of graphite and diamond to onions. *Crystals* **2018**, *8*, 68.

50. Gao, Y.; Ma, Y.; An, Q.; Levitas, V.I.; Zhang, Y.; Feng, B.; Chaudhuri, J.; Goddard, W.A. III: Shear driven formation of nano-diamonds at sub-gigapascals and 300 K. *Carbon* **2019**, *146*, 364–368.

51. Edalati, K.; Uehiro, R.; Fujiwara, K.; Ikeda, Y.; Li, H.W.; Sauvage, X.; Valiev, R.Z.; Akiba, E.; Tanaka, I.; Horita, Z. Ultra-severe plastic deformation: Evolution of microstructure, phase transformation and hardness in immiscible magnesium-based systems. *Mater. Sci. Eng. A* **2017**, *701*, 158–166.

52. Oberdorfer, B.; Lorenzoni, B.; Unger, K.; Sprengel, W.; Zehetbauer, M.; Pippan, R.; Wurschum, R. Absolute concentration of free volume-type defects in ultrafine-grained Fe prepared by high-pressure torsion. *Scr. Mater.* **2010**, *63*, 452–455.

53. Straumal, B.B.; Mazilkin, A.A.; Baretzky, B.; Schütz, G.; Rabkin, E.; Valiev, R.Z. Accelerated diffusion and phase transformations in Co-Cu alloys driven by the severe plastic deformation. *Mater. Trans.* **2012**, *53*, 63–71.

54. Edalati, K.; Emami, H.; Staykov, A.; Smith, D.J.; Akiba, E.; Horita, Z. Formation of metastable phases in magnesium-titanium system by high-pressure torsion and their hydrogen storage performance. *Acta Mater.* **2015**, *50*, 150–156.

55. Kormout, K.S.; Pippan, R.; Bachmaier, A. Deformation-induced supersaturation in immiscible material systems during high-pressure torsion. *Adv. Eng. Mater.* **2017**, *19*, 1600675.

56. Bridgman, P.W. Effects of high shearing stress combined with high hydrostatic pressure. *Phys. Rev.* **1935**, *48*, 825–847.

57. Edalati, K.; Horita, Z. Application of high-pressure torsion for consolidation of ceramic Powders. *Scr. Mater.* **2010**, *63*, 174–177.

58. Edalati, K.; Toh, S.; Ikomaa, Y.; Horita, Z. Plastic deformation and allotropic phase transformations in zirconia ceramics during high-pressure torsion. *Scr. Mater.* **2011**, *65*, 974–977.

59. Makhnev, A.A.; Nomerovannaya, L.V.; B.A.; Gizhevskii, B.A.; Naumov, S.V.; Kostromitin, N.V. Effect of redistribution of the optical spectral weight in CuO nanostructured ceramics. *Solid State Phenom.* **2011**, *168–169*, 285–288.

60. Gizhevskii, B.A.; Sukhorukov, Y.P.; Nomerovannaya, L.V.; Makhnev, A.A.; Ponosov, Y.S.; Telegin, A.V.; Mostovshchikov, E.V. Features of optical properties and the electronic structure of nanostructured oxides of 3d-Metals. *Solid State Phenom.* **2011**, *168–169*, 317–320.

61. Delogu, F. Are processing conditions similar in ball milling and high-pressure torsion? The case of the tetragonal-to-monoclinic phase transition in $ZrO_2$ powders. *Scr. Mater.* **2012**, *67*, 340–343.

62. Mostovshchikova, E.V.; Gizhevskii, B.A.; Loshkareva, N.N.; Galakhov, V.R.; Naumov, S.V.; Ponosov, Y.S.; Ovechkina, N.A.; Kostromitina, N.V.; Buling, A.; Neumann, M. Infrared and X-ray absorption spectra of $Cu_2O$ and CuO nanoceramics. *Solid State Phenom.* **2012**, *190*, 683–686.

63. Telegin, A.V.; Gizhevskii, B.A.; Nomerovannaya, L.V.; Makhnev, A.A. The optical and magneto-optical properties of nanostructured oxides of 3d-metals. *J. Supercond. Nov. Magn.* **2012**, *25*, 2683–2686.

64. Edalati, K.; Arimura, M.; Ikoma, Y.; Daio, T.; Miyata, M.; Smith, D.J.; Horita, Z. Plastic deformation of BaTiO3 ceramics by high-pressure torsion and changes in phase transformations, optical and dielectric properties. *Mater. Res. Lett.* **2015**, *3*, 216–221.

65. Razavi-Khosroshahi, H.; Edalati, K.; Arita, M.; Horita, Z.; Fuji, M. Plastic strain and grain size effect on high-pressure phase transformations in nanostructured $TiO_2$ ceramics. *Scr. Mater.* **2016**, *124*, 59–62.

66. Razavi-Khosroshahi, H.; Edalati, K.; Emami, H.; E.; Akiba, E.; Z.; Horita, Z.; Fuji, M. Optical properties of nanocrystalline monoclinic Y2O3 stabilized by grain size and plastic strain effects via high-pressure torsion. *Inorg. Chem.* **2017**, *56*, 2576–2580.

67. Kuznetsova, E.I.; Degtyarev, M.V.; Zyuzeva, N.A.; Bobylev, I.B.; Pilyugin, V.P. Microstructure of YBa2Cu3Oy subjected to severe plastic deformation by high pressure torsion. *Phys. Met. Metallogr.* **2017**, *118*, 815–823.

68. Feng, B.; Levitas, V.I. Coupled elastoplasticity and plastic strain-induced phase transformation under high pressure and large strains: Formulation and application to BN sample compressed in a diamond anvil cell. *Int. J. Plast.* **2017**, *96*, 156–181.

69. Qian, C.; He, Z.; Liang, C.; Ji, W. New in situ synthesis method for Fe3O4/flake graphite nanosheet composite structure and its application in anode materials of lithium-ion batteries. *J. Nanomater.* **2018**, *2018*, 2417949.

70. Qi, Y.; Kosinova, A.; Kilmametov, A.R.; Straumal, B.B.; Rabkin, E. Plastic flow and microstructural instabilities during high-pressure torsion of Cu/ZnO composites. *Mater. Charact.* **2018**, *145*, 389–401.

71. Shabashov, V.A.; Makarov, A.V.; Kozlov, K.A.; Sagaradze, V.V.; Zamatovskii, A.E.; Volkova, E.G.; Luchko, S.N. Deformation-induced dissolution and precipitation of nitrides in austenite and ferrite of a high-nitrogen stainless steel. *Phys. Met. Metallogr.* **2018**, *119*, 193–204.

72. Feng, B.; Levitas, V.I.; Li, W. FEM modeling of plastic flow and strain-induced phase transformation in BN under high pressure and large shear in a rotational diamond anvil cell. *Int. J. Plast.* **2019**, *113*, 236–254.

73. Edalati, K.; Wang, Q.; Eguchi, H.; Razavi-Khosroshahi, H.; Emami, H.; Yamauchi, M.; Fuji, M.; Horita, Z. Impact of TiO2-II phase stabilized in anatase matrix by high-pressure torsion on electrocatalytic hydrogen production. *Mater. Res. Lett.* **2019**, *7*, 334–339.

74. Edalati, K. Review on recent advancements in severe plastic deformation of oxides by high-pressure torsion (HPT). *Adv. Eng. Mater.* **2019**, *21*, 1800272.

75. Permyakova, I.; Glezer, A. Amorphous-nanocrystalline composites prepared by high-pressure torsion. *Metals* **2020**, *10*, 511.

76. Fujita, I.; Edalati, P.; Wang, Q.; Watanabe, M.; Arita, M.; Munetoh, S.; Ishihara, T.; Edalati, K. Novel black bismuth oxide (Bi2O3) with enhanced photocurrent generation, produced by high-pressure torsion straining. *Scr. Mater.* **2020**, *187*, 366–370.

77. Wang, Q.; Watanabe, M.; Edalati, K. Visible-light photocurrent in nanostructured high-pressure TiO2-II (Columbite) phase. *J. Phys. Chem. C* **2020**, *124*, 13930–13935.

78. Edalati, K.; Fujita, I.; Sauvage, X.; Arita, M.; Horita, Z. Microstructure and phase transformations of silica glass and vanadium oxide by severe plastic deformation via high-pressure torsion straining. *J. Alloys Compd.* **2019**, *779*, 394–398.

79. Edalati, K.; Wang, Q.; Razavi-Khosroshahi, H.; Emami, H.; Fuji, M.; Horita, Z. Low-temperature anatase-to-rutile phase transformation and unusual grain coarsening in titanium oxide nanopowders by high-pressure torsion straining. *Scr. Mater.* **2019**, *162*, 341–344.

80. Qi, Y.; Kauffmann, Y.; Kosinova, A.; Kilmametov, A.R.; Straumal, B.B.; Rabkin, E. Gradient bandgap narrowing in severely deformed ZnO nanoparticles. *Matter. Res. Lett.* **2021**, *9*, 58–64.

81. Akrami, S.; Edalati, P.; Edalati, K.; Fuji, M. High-entropy ceramics: A review of principles, production and applications. *Mater. Sci. Eng. R* **2021**, *146*, 100644.

82. Oses, C.; Toher, C.; Curtarolo, S. High-entropy ceramics. *Nat. Rev. Mater.* **2020**, *5*, 295–309.

83. Sarkar, A.; Velasco, L.; Wang, D.; Wang, Q.; Talasila, G.; de Biasi, L.; Kübel, C.; Brezesinski, T.; Bhattacharya, S.S.; Hahn, H.; et al. High entropy oxides for reversible energy storage. *Nat. Commun.* **2018**, *9*, 3400.

84. Chen, H.; Lin, W.; Zhang, Z.; Jie, K.; Mullins, D.R.; Sang, X.; Yang, S.Z.; Jafta, C.J.; Bridges, C.A.; Hu, X. Mechanochemical synthesis of high entropy oxide materials under ambient conditions: Dispersion of catalysts via entropy maximization. *ACS, Mater. Lett.* **2019**, *1*, 83–88.

85. Radon, A.; Hawełek, D.; Łukowiec, J.; Kubacki, P.; Włodarczyk, P. Dielectric and electromagnetic interference shielding properties of high entropy (Zn, Fe, Ni, Mg, Cd)Fe₂O₄ ferrite. *Sci. Rep.* **2019**, *9*, 20078.

86. Witte, R.; Sarkar, A.; Kruk, R.; Eggert, B.; Brand, R.A.; Wende, H.; Hahn, H. High entropy oxides: An emerging prospect for magnetic rare-earth transition metal perovskites. *Phys. Rev. Mater.* **2019**, *3*, 034406.

87. Wright, A.J.; Huang, C.; Walock, M.J.; Ghoshal, A.; Murugan, M.; Luo, J. Sand corrosion, thermal expansion, and ablation of medium-and high-entropy compositionally complex fluorite oxides. *J. Am. Ceram. Soc.* **2021**, *104*, 448–462.

88. Takata, T.; Pan, C.; Domen, K. Recent progress in oxynitride photocatalysts for visible-light-driven water splitting. *Sci. Technol. Adv. Mater.* **2015**, *16*, 033506.

89. Wang, K.; Lu, J.; Lu, y.; Lau, C.H.; Zheng, Y.; Fan, X. Unravelling the C-C coupling in CO₂ photocatalytic reduction with H₂O on Au/TiO₂₋ₓ: Combination of plasmonic excitation and oxygen vacancy. *Appl. Catal. B* **2021**, *292*, 120147.

90. Lu, H.; Tournet, J.; Dastafkan, K.; Liu, Y.; Ng, Y.H.; Karuturi, S.K.; Zhao, C.; Yin, Z. No-blemetal- free multicomponent nanointegration for sustainable energy conversion. *Chem. Rev.* **2021**, *121*, 10271–10366.

91. Wang, Y.; Chen, Y.; Zuo, Y.; Wang, F.; Yao, J.; Li, B.; Kang, S.; Li, X.; Cui, L. Hierarchically mesostructured TiO₂/graphitic carbon composite as a new efficient photocatalyst for the reduction of CO₂ under simulated solar irradiation. *Catal. Sci. Technol.* **2013**, *3*, 3286–3291.

92. Liu, L.J; Pitts, D.T.; Zhao, H.L.; Zhao, C.Y.; Li, Y. Silver-incorporated bicrystalline (anatase/brookite) TiO₂ microspheres for CO₂ photoreduction with water in the presence of methanol. *Appl. Catal. A* **2013**, *467*, 474–482.

93. Sekizawa, K.; Maeda, K.; Domen, K.; Koike, K.; Ishitani, O. Artificial Z-scheme constructed with a supramolecular metal complex and semiconductor for the photocatalytic reduction of CO₂. *J. Am. Chem. Soc.* **2013**, *135*, 4596–4599.

94. Tahir, M.; Amin, N.S. Photocatalytic reduction of carbon dioxide with water vapors over montmorillonite modified TiO₂ nanocomposites. *Appl. Catal. B* **2013**, *142–143*, 512–522.

95. Zhang, Z.Y.; Wang, Z.; Cao, S.W.; Xue, C. Au/Pt nanoparticle-decorated TiO₂ nanofibers with plasmon-enhanced photocatalytic activities for solar-to-fuel conversion. *J. Phys. Chem. C* **2013**, *117*, 25939–25947.

96. He, Z.; Wen, L.; Wang, D.; Xue, Y.; Lu, Q.; Wu, C.; Chen, J.; Song, S. Photocatalytic reduction of CO₂ in aqueous solution on surface-fuorinated anatase TiO₂ nanosheets with exposed {001} fcets. *Energy Fuels* **2014**, *28*, 3982–3993.

97. Fang, B.Z.; Bonakdarpour, A; Reilly, K.; Xing, Y.L.; Taghipour, F; Wilkinson, D.P. Large-scale synthesis of TiO₂ microspheres with hierarchical nanostructure for highly efficient photodriven reduction of CO₂ to CH₄. *ACS Appl. Mater. Interfaces* **2014**, *6*, 15488–15498.

98. Li, K.; Lin, L.; Peng, T.; Guo, Y.; Li, R.; Zhang, J. Asymmetric zinc porphyrin-sensitized nanosized TiO₂ for efficient visible-light-driven CO₂ photoreduction to CO/CH₄. *J. Chem. Commun.* **2015**, *51*, 12443–12446.

99. Fang, B.Z.; Xing, Y.L.; Bonakdarpour, A.; Zhang, S.C; Wilkinson, D.P. Hierarchical CuO-TiO₂ hollow microspheres for highly efficient photodriven reduction of CO₂ to CH₄. *ACS Sustain. Chem. Eng.* **2015**, *3*, 2381–2388.

100. Tahir, M.; Amin, N.S. Indium-doped TiO₂ nanoparticles for photocatalytic CO₂ reduction with H₂O vapors to CH₄. *Appl. Catal. B* **2015**, *162*, 98–109.

101. Xiong, Z.; Wang, H.B.; Xu, N.Y.; Li, H.L.; Fang, B.Z.; Zhao, Y.C.; Zhang, J.Y.; Zheng, C.G. Photocatalytic reduction of CO₂ on Pt²⁺–Pt⁰/TiO₂ nanoparticles under UV/Vis light irradiation: A combination of Pt2+ doping and Pt nanoparticles deposition. *Int. J. Hydrog. Energy* **2015**, *40*, 10049–10062.

102. Ye, L.; Wang, H.; Jin, X.; Su, Y.; Wang, D.; Xie, H.; Liu, X.; Liu, X. Synthesis of olive- green few-layered BiOI for efficient photoreduction of CO₂ into solar fuels under visible/near-infrared light. *Sol. Energy Mater. Sol. Cells* **2016**, *144*, 732–739.

103. Nakada, A.; Nakashima, T.; Sekizawa, K.; Maeda, K.; Ishitani, O. Visible-light-driven CO₂ reduction on a hybrid photocatalyst consisting of a Ru(II) binuclear complex and a Ag-loaded TaON in aqueous solutions. *Chem. Sci.* **2016**, *7*, 4364–4371.

104. Ye, T.; Huang, W.; Zeng, L.; Li, M.; Shi, J. CeO2-x platelet from monometallic cerium layered double hydroxides and its photocatalytic reduction of CO₂. *Appl. Catal. B* **2017**, *210*, 141–148.

105. Pastor, E.; Pesci, F.; Reynal, A.; Handoko, A.; Guo, M.; Jan, X.; Cowan, A.; Klug, D.; Durrant, J.; Tang. J. Interfacial charge separation in Cu₂O/RuOₓ as a visible light driven CO₂ reduction catalyst. *Phys. Chem. Chem. Phys.* **2014**, *16*, 5922–5926.

106. Jiao, J.; Wei, Y.; Zhao, Y.; Zhao, Z.; Duan, A.; Liu, J.; Pang, Y.; Li, J.; Jiang, G.; Wang, Y. AuPd/3DOM-TiO₂ catalysts for photocatalytic reduction of CO₂: High efficient separation of photogenerated charge carriers. *Appl. Catal. B* **2017**, *209*, 228–239.

107. Kim, C.; Cho, K.M.; Al-Saggaf, A.; Gereige, I.; Jung, H.T. Z-scheme photocatalytic CO₂ conversion on three-dimensional BiVO₄/carbon-coated Cu₂O nanowire arrays under visible light. *ACS Catal.* **2018**, *8*, 4170–4177.

108. Wang, J.; Qin, C.; Wang, H.; Chu, M.; Zada, A.; Zhang, X.; Li, J.; Raziq, F.; Qu, Y.; Jing, L. Exceptional photocatalytic activities for CO₂ conversion on AlO bridged g-C₃N₄/α-Fe₂O₃ z-scheme nanocomposites and mechanism insight with isotopes Z. *Appl. Catal. B* **2018**, *224*, 459–466.

109. Jiang, H.; Katsumata, K.; Hong, J.; Yamaguchi, A.; Nakata, K.; Terashima, C.; Matsushita, N.; Miyauchi, M.; Fujishima, A. Photocatalytic reduction of CO₂ on Cu₂O-loaded Zn-Cr layered double hydroxides. *Appl. Catal. B* **2018**, *224*, 783–790.

110. Crake, A.; Christoforidis, K.C.; Godin, R.; Moss, B.; Kafizas, A.; S.; Zafeiratos, S.; Durrant, J.R.; Petit, C. Titanium dioxide/carbon nitride nanosheet nanocomposites for gas phase CO₂ photoreduction under UV-visible irradiation. *Appl. Catal. B* **2019**, *242*, 369–378.

111. Li, Y.; Wang, C.; Song, M.; Li, D.; Zhang, X.; Liu, Y. TiO₂₋ₓ/CoOₓ Photocatalyst sparkles in photothermocatalytic reduction of CO₂ with H₂O steam. *Appl. Catal. B* **2019**, *243*, 760–770.

112. Bai, Y.; Yang, P.; Wang, L.; Yang, B.; Xie, H.; Zhou, Y.; Ye, L. Ultrathin Bi₄O₅Br₂ nanosheets for selective photocatalytic CO₂ conversion into CO. *Chem. Eng. J.* **2019**, *360*, 473–482.

113. Maiti, D.; Meier, A.J.; Cairns, J.; Ramani, S.; Martinet, K.; Kuhn, J.N.; Bhethanabotla, V.R. Intrinsically strained noble metal-free oxynitrides for solar photoreduction of CO₂. *Dalton Trans.* **2019**, *48*, 12738–12748.

114. Kozlova, E.A.; Lyulyukin, M.N.; Markovskaya, D.V.; Selishchev, D.S.; Cherepanova, S.V.; Kozlov, D.V.; Synthesis of Cd₁₋ₓZnₓS photocatalysts for gas-phase CO₂ reduction under visible light. *Photochem. Photobiol. Sci.* **2019**, *18*, 871–877.

115. Jin, X.; Lv, C.; Zhou, X.; Ye, L.; Xie, H.; Liu, Y.; Su, H.; Zhang, B.; Chen, G. Oxygen vacancies engineering Bi₂₄O₃₁Cl₁₀ photocatalyst for boosted CO₂ conversion. *ChemSusChem* **2019**, *12*, 2740–2747.

116. Guo, S.; Di, J.; Chen, C.; Zhu, C.; Duan, M.; Lian, C.; Ji, M.; Zhou, W.; Xu, M.; Song, P.; et al. Oxygen vacancy mediated bismuth stannate ultra-small nanoparticle towards photocatalytic CO₂-to-CO conversion. *Appl. Catal. B* **2020**, *276*, 119156.

117. Duan, Z.; Zhao, X.; Wei, C.; Chen, L. Ag-Bi/BiVO₄ chain-like hollow microstructures with enhanced photocatalytic activity for CO₂ conversion. *Appl. Catal. A* **2020**, *594*, 117459.

118. Chen, Y.; Wang, F.; Cao, Y.; Zhang, F.; Zou, Y.; Huang, Z.; Ye, L.; Zhou, Y. Interfacial oxygen vacancy engineered two-dimensional g-C₃N₄/BiOCl heterostructures with boosted photocatalytic conversion of CO₂. *ACS Appl. Energy Mater.* **2020**, *3*, 4610–4618.

119. Dao, X.Y.; Xie, X.F.; Guo, J.H.; Zhang, X.Y.; Kang, Y.S.; Sun, W.Y. Boosting photocatalytic CO₂ reduction efficiency by heterostructures of NH₂-MIL-101 (Fe)/g-C₃N₄. *ACS Appl. Energy Mater.* **2020**, *3*, 3946–3954.

120. Zhang, X.; Ren, G.; Zhang, C.; Li, R.; Zhao, Q.; Fan, C. Photocatalytic reduction of CO₂ to CO over 3D Bi₂MoO₆ microspheres: Simple synthesis, high efficiency and selectivity, reaction mechanism. *Catal. Lett.* **2020**, *150*, 2510–2516.

121. Xie, Y.; Zhuo, Y.; Liu, Lin, Y.; Zuo, D.; Wu, X.; Li, C.; Wong, P.K. Ternary g-C₃N₄/ZnNCN@ZIF-8 hybrid photocatalysts with robust interfacial interactions and enhanced CO₂ reduction, *Solar RRL.* **2020**, *4*, 1900440.

122. Lin, N.; Lin, Y.; Qian, X.; Wang, X.; Su, W. Construction of a 2D/2D WO₃/LaTiO₂N direct Z-scheme photocatalyst for enhanced CO₂ reduction performance under visible light. *ACS Sustain. Chem. Eng.* **2021**, *9*, 13686–1369.

123. Song, J.; Lu, Y.; Lin, Y.; Liu, Q.; Wang, X.; Su, W. A direct Z-scheme α-Fe₂O₃/LaTiO₂N visible-light photocatalyst for enhanced CO₂ reduction activity. *Appl. Catal. B* **2021**, *292*, 120185.

124. Valiev, R.Z.; Islamgaliev, R.K.; Alexandrov, I.V. Bulk nanostructured materials from severe plastic deformation. *Prog. Mater. Sci.* **2000**, *45*, 103–189.

125. Valiev, R.Z.; Estrin, Y.; Horita, Z.; Langdon, T.G.; Zehetbauer, M.J.; Zhu, Y.T. Producing bulk ultrafine-grained materials by severe plastic deformation. *JOM* **2006**, *58*, 33–39.

126. Azushima, A.; Kopp, R.; Korhonen, A.; Yang, D.Y.; Micari, F.; Lahoti, G.D.; Groche, P.; Yanagimoto, J.; Tsuji, N.; Rosochowski, A.; et al. Severe plastic deformation (SPD) processes for metals. *CRIP Ann. Manuf. Technol.* **2008**, *57*, 716–735.

127. Segal, V. Review: Modes and processes of severe plastic deformation (SPD). *Materials*, **2018**, *11*, 1175.

128. Zehetbauer, M.; Grossinger, R.; Krenn, H.; Krystian, M.; Pippan, R.; Rogl, P.; Waitz, T.; Wurschum, R. Bulk nanostructured functional `materials by severe plastic deformation. *Adv. Eng. Mater.* **2010**, *12*, 692–700.

129. Estrin, Y.; Vinogradov., A. Extreme grain refinement by severe plastic deformation: A wealth of challenging science. *Acta Mater.* **2013**, *61*, 782–817.

130. Edalati, K.; Horita, Z. Special issue on severe plastic deformation for nanomaterials with advanced functionality. *Mater. Trans.* **2019**, *60*, 1103.

131. Bryła, K.; Edalati, K. Historical studies by polish scientist on ultrafine-grained materials by severe plastic deformation. *Mater. Trans.* **2019**, *60*, 1553–1560.

132. Horita, Z.; Edalati, K. Severe plastic deformation for nanostructure controls. *Mater. Trans.* **2020**, *61*, 2241–2247.

133. Pereira, P.H.R.; Figueiredo, R.B. Finite element modelling of high-pressure torsion: An overview. *Mater. Trans.* **2019**, *60*, 1139–1150.

134. Levitas, V.I. High-pressure phase transformations under severe plastic deformation by torsion in rotational anvils. *Mater. Trans.* **2019**, *60*, 1294–1301.

135. Tsuji, N.; Gholizadeh, R.; Ueji, R.; Kamikawa, N.; Zhao, L.; Tian, Y.; Bai, Y.; Shibata, A. Formation mechanism of ultrafine grained microstructures: Various possibilities for fabricating bulk nanostructured metals and alloys. *Mater. Trans.* **2019**, *60*, 1518–1532.

136. Renk, O.; Pippan, R. Saturation of grain refinement during severe plastic deformation of single phase materials: Reconsiderations, current status and open questions. *Mater. Trans.* **2019**, *60*, 1270–1282.

137. Popov, V.V.; Popova, E.N. Behavior of Nb and Cu¬Nb composites under severe plastic deformation and annealing. *Mater. Trans.* **2019**, *60*, 1209–1220.

138. Skrotzki, W. Deformation heterogeneities in equal channel angular pressing. *Mater. Trans.* **2019**, *60*, 1331–1343.

139. Miura, H.; Iwama, Y.; Kobayashi, M. Comparisons of microstructures and mechanical properties of heterogeneous nanostructure induced by heavy cold rolling and ultrafine-grained structure by multi-directional forging of Cu¬Al alloy. *Mater. Trans.* **2019**, *60*, 1111–1115.

140. Faraji, G.; Torabzadeh, H. An overview on the continuous severe plastic deformation methods. *Mater. Trans.* **2019**, *60*, 1316–1330.

141. Toth, L.S.; Chen, C.; Pougis, A.; Arzaghi, M.; Fundenberger, J.J.; Massion, R.; Suwas, S. High pressure tube twisting for producing ultra fine grained materials: A review. *Mater. Trans.* **2019**, *60*, 1177–1191.

142. Masuda, T.; Horita, Z. Grain refinement of AZ31 and AZ61 Mg alloys through room temperature processing by up-scaled high-pressure torsion. *Mater. Trans.* **2019**, *60*, 1104–1110.

143. Yang, X.; Pan, H.; Zhang, J.; Gao, H.; Shu, B.; Gong, Y.; Zhu, X. Progress in mechanical properties of gradient structured metallic materials induced by surface mechanical attrition treatment. *Mater. Trans.* **2019**, *60*, 1543–1552.

144. Grosdidier, T.; Novelli, M. Recent developments in the application of surface mechanical attrition treatments for improved gradient structures: Processing parameters and surface reactivity. *Mater. Trans.* **2019**, *60*, 1344–1355.

145. Suwas, S.; Mondal, S. Texture evolution in severe plastic deformation processes. *Mater. Trans.* **2019**, *60*, 1457–1471.

146. Sauvage, X.; Duchaussoy, A.; Zaher, G. Strain induced segregations in severely deformed materials. *Mater. Trans.* **2019**, *60*, 1151–1158.

147. Wilde, G.; Divinski, S. Grain boundaries and diffusion phenomena in severely deformed materials. *Mater. Trans.* **2019**, *60*, 1302–1315.

148. Gubicza, J. Lattice defects and their influence on the mechanical properties of bulk materials processed by severe plastic deformation. *Mater. Trans.* **2019**, *60*, 1230–1242.

149. Čížek, J.; Janeček, M.; Vlasák, T.; Smola, B.; Melikhova, O.; Islamgaliev, R.K.; Dobatkin, S.V. The development of vacancies during severe plastic deformation. *Mater. Trans.* **2019**, *60*, 1533–1542.

150. Kunimine, T.; Watanabe, M. A comparative study of hardness in nanostructured Cu-Zn, Cu-Si and Cu-Ni solid-solution alloys processed by severe plastic deformation. *Mater. Trans.* **2019**, *60*, 1484–1488.

151. Kuramoto, S.; Furuta, T. Severe plastic deformation to achieve high strength and high ductility in Fe¬Ni based alloys with lattice softening. *Mater. Trans.* **2019**, *60*, 1116–1122.

152. Kawasaki, M.; Langdon, T.G. The contribution of severe plastic deformation to research on superplasticity. *Mater. Trans.* **2019**, *60*, 1123–1130.

153. Demirtas, M.; Purcek, G. Room temperature superplaticity in fine/ultrafine grained materials subjected to severe plastic deformation. *Mater. Trans.* **2019**, *60*, 1159–1167.

154. Moreno-Valle, E.C.; Pachla, W.; Kulczyk, M.; Sabirov, I.; Hohenwarter, A. Anisotropy of tensile and fracture behavior of pure titanium after hydrostatic extrusion. *Mater. Trans.* **2019**, *60*, 2160–2167.

155. Kral, P.; Dvorak, J.; Sklenicka, V.; Langdon, T.G. The characteristics of creep in metallic materials processed by severe plastic deformation. *Mater. Trans.* **2019**, *60*, 1506–1517.

156. Razavi-Khosroshahi, H.; Fujii M. Development of metal oxide high-pressure phases for photocatalytic properties by severe plastic deformation. *Mater. Trans.* **2019**, *60*, 1203–1208.

157. Ikoma, Y. Severe plastic deformation of semiconductor materials using high-pressure torsion. *Mater. Trans.* **2019**, *60*, 1168–1176.

158. Blank, V.D.; Popov, M.Y.; Kulnitskiy, B.A. The effect of severe plastic deformations on phase transitions and structure of solids. *Mater. Trans.* **2019**, *60*, 1500–1505.

159. Révész, Á.; Kovács, Z. Severe plastic deformation of amorphous alloys. *Mater. Trans.* **2019**, *60*, 1283–1293.

160. Beloshenko, V.; Vozniak, I.; Beygelzimer, Y.; Estrin, Y.; Kulagin, R. Severe plastic deformation of polymers. *Mater. Trans.* **2019**, *60*, 1192–1202.

161. Mazilkin, A.; Straumal, B.; Kilmametov, A.; Straumal, P.; Baretzky, B. Phase transformations induced by severe plastic deformation. *Mater. Trans.* **2019**, *60*, 1489–1499.

162. Bachmaier, A.; Pippan, R. High-pressure torsion deformation induced phase transformations and formations: New material combinations and advanced properties. *Mater. Trans.* **2019**, *60*, 1256–1269.

163. Han, J.K.; Jang, J.I.; Langdon, T.G.; Kawasaki, M. Bulk-state reactions and improving the mechanical properties of metals through high-pressure torsion. *Mater. Trans.* **2019**, *60*, 1131–1138.

164. Edalati, K. Metallurgical alchemy by ultra-severe plastic deformation via high-pressure torsion process. *Mater. Trans.* **2019**, *60*, 1221–1229.

165. Mito, M.; Shigeoka, S.; Kondo, H.; Noumi, N.; Kitamura, Y.; Irie, K.; Nakamura, K.; Takagi, S.; Deguchi, H.; Tajiri, T.; et al. Hydrostatic compression effects on fifth-group element superconductors V, Nb, and Ta subjected to high-pressure torsion. *Mater. Trans.* **2019**, *60*, 1472–1483.

166. Nishizaki, T.; Edalati, K.; Lee, S.; Horita, Z.; Akune, T.; Nojima, T.; Iguchi, S.; Sasaki, T. Critical temperature in bulk ultrafine-grained superconductors of Nb, V, and Ta processed by high-pressure torsion. *Mater. Trans.* **2019**, *60*, 1367–1376.

167. Rogl, G.; Zehetbauer, M.J.; Rogl, P.F. The effect of severe plastic deformation on thermoelectric performance of skutterudites, half-Heuslers and Bi-tellurides. *Mater. Trans.* **2019**, *60*, 2071–2085.
168. Enikeev, N.A.; Shamardin, V.K.; Radiguet, B. Radiation tolerance of ultrafine-grained materials fabricated by severe plastic deformation. *Mater. Trans.* **2019**, *60*, 1723–1731.
169. Leiva, D.R.; Jorge, A.M.; Ishikawa Jr., T.T.; Botta, W.J. Hydrogen storage in Mg and Mg-based alloys and composites processed by severe plastic deformation. *Mater. Trans.* **2019**, *60*, 1561–1570.
170. Huot, J.; Tousignant, M. Effect of cold rolling on metal hydrides. *Mater. Trans.* **2019**, *60*, 1571–1576.
171. Miyamoto, H.; Yuasa, M.; Rifai, M.; Fujiwara, H. Corrosion behavior of severely deformed pure and single-phase materials. *Mater. Trans.* **2019**, *60*, 1243–1255.
172. Valiev, R.Z.; Parfenov, E.V.; Parfenova, L.V. Developing nanostructured metals for manufacturing of medical implants with improved design and biofunctionality. *Mater. Trans.* **2019**, *60*, 1356–1366.
173. Lowe, T.C.; Valiev, R.Z.; Li, X.; Ewing, B.R. Commercialization of bulk nanostructured metals and alloys. *MRS Bull.* **2021**, *46*, 265–272.